\documentstyle[aps,pre,multicol,epsf]{revtex}
\def\be{\begin{equation}}
\def\ee{\end{equation}}

\def\bea{\begin{eqnarray}}
\def\eea{\end{eqnarray}}
\begin{document}
\draft
\title{Formation of Liesegang patterns: A spinodal decomposition
scenario}
\vspace {1truecm}
\author{T. Antal${}^1$, M. Droz${}^1$, J. Magnin${}^1$,
and Z. R\'acz${}^{2}$}
\address{{}$^1$ {D\'epartement de Physique Th\'eorique, Universit\'e de
Gen\`eve, CH 1211 Gen\`eve 4, Switzerland.}}
\address{{}$^2$Institute for Theoretical Physics,
E\"otv\"os University, 1117 Budapest, P\'azm\'any s\'et\'any 1/a, Hungary}

\date{\today}

\maketitle

\begin{abstract}
Spinodal decomposition in the presence of a moving particle 
source is proposed as a mechanism for the formation of Liesegang bands. 
This mechanism yields a sequence of band positions $x_n$ that
obeys the spacing law $x_n\sim Q(1+p)^n$. The dependence of the parameters  
$p$ and $Q$ on the initial concentration of the reagents 
is determined and we find that the functional form of $p$ is
in agreement with the experimentally observed Matalon-Packter law. 
\end{abstract}

\date{\today}
\begin{multicols}{2}
\narrowtext

Pattern-forming chemical, physical and biological processes are common
in nature and  patterns often emerge in the wake of a moving front
\cite{Hoh}.  In particular, when an electrolyte $A$ diffuses into a gel
containing another electrolyte $B$, the eventual formation of a rhythmic
pattern of precipitate by the moving chemical reaction front is known
as the {\it Liesegang phenomenon}~\cite{liese,Henisch}.   
The observed precipitation patterns usually consist of
a set of bands or rings (depending on the geometry of the system)
clearly separated in the direction perpendicular to the motion of the front. 
This phenomenon is believed to be responsible for many precipitation patterns
such as e.g. the structure of agate rocks \cite{Henisch}.
Although the Liesegang phenomenon has been studied for a century, 
the mechanisms responsible for these structures is still
under discussion \cite{us}. 

Most of the reproducible Liesegang patterns 
are characterizable by the following generic laws. First, 
the position of the $n$-th band $x_n$
(measured from the initial interface of the reagents) is proportional to 
$\sqrt{t_n}$ where $t_n$ is the time elapsed till the appearance of the band. 
This so-called {\it time law}~\cite{morse} 
is a direct consequence of the diffusive nature of the dynamics.
Secondly, the positions $x_n$ of the
bands usually form a geometric series
({\it spacing law}~\cite{jabli}):
\begin{equation}
x_{n} 
\stackrel{{\rm n \, large}}{-\hskip -4.6pt -\hskip -4.7pt \longrightarrow} 
Q(1+p)^n 
\label{spacing}
\end{equation}
where $p>0$ is called the spacing coefficient and $Q$ is the 
amplitude of the spacing law.
Finally, the width $w_n$ of the bands have been observed to
increase with $n$ and to obey the {\it width law}~\cite{widthlaw},
$ w_n \sim x_n$.

Most of the detailed experimental observations concern the spacing law. It 
has been found that the spacing coefficient $p$ is a nonuniversal quantity 
depending (among other parameters) on the experimentally controllable 
concentrations $a_0$ and $b_0$ of 
the outer ($A$) and  inner ($B$) electrolytes.
This dependence is expressed by the 
{\it Matalon-Packter law}~\cite{{Matalon},{Packter}}:
\begin{equation}
p=F(b_0) + G(b_0)\frac{b_0}{a_0} \quad ,
\label{MatPac}
\end{equation}
where $F$ and $G$ are decreasing functions of their 
argument $b_0$ \cite{matpremark}.

The task of theories is to explain the existence of patterns 
obeying the spacing law (\ref{spacing}) and, on a more sophisticated level,
to derive the Matalon-Packter law (\ref{MatPac}). 
The theoretical approaches proposed up to now can be divided into 
two categories. 
The first one contains the {\it ion-product supersaturation}
type theories~\cite{Wagner,Prager,zeldo} where the outer
and inner reactants turn directly into precipitate ($A+B\to D$)
whenever their local 
concentration product is above a threshold $q^*$.
The second category contains the {\em intermediate-compound} 
theories~\cite{dee,luthi} which assume 
the existence of a species $C$ ($A+B\to C\to D$).
The $C$-s are produced in a reaction-diffusion front 
that moves through the system and the precipitation  $C\to D$ takes place 
only if the local concentration $c$ reaches some threshold value $c^*$. 
In the presence of $D$, the process continues until the 
concentration of $C$ drops below another threshold $d^*$. 
The problem with these theories is that either they 
contain parameters such as e.g. $d^*$
that are difficult to control experimentally and 
not easy to grasp theoretically, or they describe a detailed 
mechanism that is too complicated to deduce quantitative consequences 
such as the Matalon-Packter law. 
Our goal with this Letter is to describe the 
formation of Liesegang patterns in terms of a model that
contains the basic mechanism of phase separation that underlies the
band formation but, at the same time, simple 
enough so that quantitative predictions are readily made. 
In particular we show the existence of the spacing law, obtain the 
Matalon-Packter law with estimates of $F(b_0)$ and $G(b_0)$ and,
furthermore, the $b_0/a_0$ dependence of the amplitude $Q$ of
the spacing law is also determined.

All the theories discussed above can produce the spacing law but the 
comparison with the Matalon-Packter law suggest that the 
intermediate-compound theories are clearly preferable \cite{us}. Accordingly, 
we shall assume that the motion of the reaction-diffusion front and 
the dynamics of the $C$ particles deposited 
by the front should be the main ingredients of a theory
of Liesegang patterns. The dynamics of the $A+B\to C$ front 
has been solved so its motion and the production of $C$-s
are well known \cite{galfi}. Thus the new aspect of our theory is the 
model for the dynamics of $C$-s.

In our picture, the $C$-s are particles that can move only 
by diffusion due to the presence of the gel. There is an attractive interaction 
between the particles that induces aggregation at low enough temperatures and
high enough densities. An easily understandable, spatially discretized 
version of the model is the following
spin-$1/2$ kinetic Ising model with competing spin-flip 
and spin-exchange dynamics \cite{DRS}. 
Empty and
occupied lattice sites are associated with down and up spins, respectively. 
The initial state is empty (all spins are down) and
the moving front flips the down spins at a given rate (Glauber
dynamics~\cite{glauber}). The diffusion is described by a spin
exchange process (Kawasaki dynamics~\cite{kawa}). The rates of exchanges
are governed by a heat bath at temperature $T$ 
with ferromagnetic couplings between the spins modeling 
the attraction among the $C$'s. 
The experimentally observed freezing of the emerging
patterns implies that the corresponding dynamics takes place at a very low
effective temperature. If this model represents
the pattern forming process correctly then one expects the emergence 
of bands of up and down spins in the wake of the moving spin-flip front. 
In order to understand how the bands arise, let us consider the phase diagram 
of an Ising model depicted on Fig.1.
\begin{figure}[htb]
\centerline{
        \epsfxsize=8cm
        \epsfbox{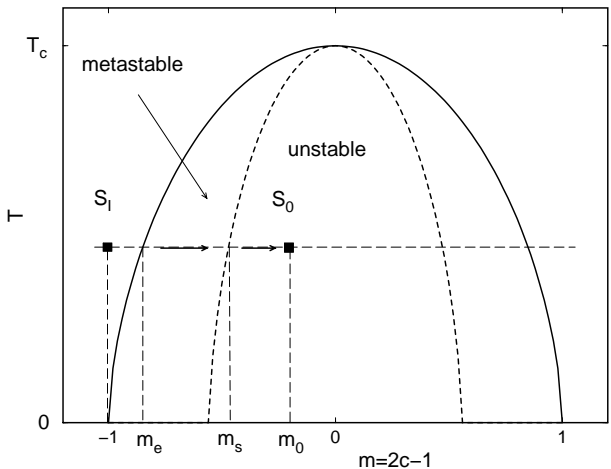}
           }
\vspace{0.5cm}
\caption{Qualitative phase diagram for the Ising model.
The magnetization $m$ is related to the density $c$ of $C$ particles 
through $m=2c-1$. The solid line is the
coexistence curve and the dotted one is the spinodal line. $S_I$ is the 
initial state with $m=-1$, $\pm m_e$ are the equilibrium magnetizations 
at a given temperature $T$ while $\pm m_s$ are the magnetizations at the 
spinodal line. The front alone would leave behind a magnetization $m_0$.}
\label{Fig1}
\end{figure}

One starts from the all-spins-down state (empty state, $S_I$ in Fig.1). 
Since an $A+B\to C$ reaction front leaves behind a constant density $c_0$
of $C$-s \cite{galfi}, the spin-flip front is 
chosen such that it produces a local magnetization $m_0=2c_0-1$. 
We would like the front to bring the system into the unstable 
regime, thus $m_0>m_s$ is assumed.
As $m$ is increasing from $-1$ to $m_0$ in the front , 
the local state moves from $S_I$ towards $S_0$ (Fig.1). 
With time, one crosses the coexistence line ($m=m_e$) and enters 
into the metastable
phase~\cite{gun}. Small clusters of up-spins 
nucleate at and aggregate behind the front. However,  
the nucleation is an activated process and so
its characteristic time-scale $\tau_{nucl}$ is large at low temperatures. 
If $\tau_{nucl}$ is much larger than the time $\tau_{front}$ needed by the front 
to put out the local magnetization $m_0$ then the system reaches the 
unstable state i.e. crosses the spinodal line ($m=m_s$) \cite{spinodal}.
Once the spinodal line is crossed, the phase separation takes place on a
short time-scale and a spin-up domain is rapidly 
formed at or behind the front, hence the formation of a 
Liesegang band. 

This band acts as a sink for the up-spins and, in the 
vicinity of the band, the local
magnetization decreases and the front
is no longer in the unstable region of the phase space.
When the front moves far enough, the effect of the band as a sink diminishes.
Thus the magnetization grows and the spinodal line is crossed 
again resulting in the formation of the next band. The repetition of this 
process should lead then to the Liesegang pattern.
The new feature of the picture described above is the assumption that
the state of the front is quasiperiodically driven into the 
{\em unstable} regime. 

The above microscopic picture can be described on the mesoscopic level as
follows. The diffusive dynamics of the coarse grained magnetization $m(x,t)$ is 
described by Model B of critical dynamics \cite{HalpHoh} with the moving 
spin-flip front appearing as a time-dependent source term $S(x,t)$: 
\bea
\partial_t m(x,t)=& 
-&\lambda \partial_x^2
\Big[\epsilon m(x,t)-\gamma m^3(x,t) +\sigma 
\partial^2_x m(x,t)\Big] \nonumber \\ 
&+& S(x,t) \label{move}
\eea
where $\lambda$ is a kinetic coefficient, $\epsilon$ measures the 
deviation from the critical temperature $T_c$ and $\epsilon>0$ 
ensures that $T<T_c$. The parameter $\gamma$ is positive to
guarantee overall stability and  
$\sigma>0$ provides the stability against short-wavelength 
fluctuations. In principle, equation 
(\ref{move}) should contain two noise terms. First, there should be 
thermal noise that, in the absence of the source, would bring the system 
to equilibrium described by the phase diagram in Fig.1. Second, there 
should be noise in the source $S$ since the origin of this
term is a reaction-diffusion process. We shall omit both noise terms. 
The thermal noise is neglected since
the effective temperature is expected to be low as discussed above. 
As to the source term, it has been shown that the properties of the 
$A+B\to C$ type reaction fronts are mean-field like above dimension two 
\cite{cornell}. We take this as an indication that the noise in $S$ can 
be neglected.

In order to complete the description of the model, we need to discuss
the actual form of the source term, $S$. The problem of the $A+B\to C$ 
reaction-diffusion front 
is well understood at the mean-field level \cite{galfi}. The source 
term is the production rate of the $C$ particles that can be calculated 
analytically. Provided 
the reagents $A$ and $B$ are spatially separated at the initial moment 
[i.e. their densities are given by 
$a(x<0,t=0)=a_0$, $b(x<0,t=0)=0$ and $a(x>0,t=0)=0$, $b(x>0,t=0)=b_0$],
the source term is a gaussian to an excellent accuracy:
\be
S(x,t)=\frac{\cal A}{t^{2/3}}\exp\Big[-\frac{[x-x_f(t)]^2}{2w^2(t)} \Big] .
\ee
The center of the front moves as $x_f(t)=\sqrt{2D_ft}$ with the 
diffusion constant $D_f$ given by the following equation
${\rm erf}(\sqrt{D_f/2D})=(a_0-b_0)/(a_0+b_0)$ with $D=D_a=D_b$ being the 
diffusion coefficient of the $A$ and $B$ particles \cite{{galfi},{koza}}.

The front is well localized since its width increases with a small power of time
$w(t)=2\sqrt{D}t^{1/6}/(ka_0K)^{1/3}$ where $k$ is the reaction rate of the 
$A+B\to C$ process, and $K$ is given by 
$K=(1+b_0/a_0)(2\sqrt{\pi})^{-1}\exp{(-D_f/D)}$.
Finally, the amplitude of the source can be expressed as
${\cal A}=0.3 k a_0^2 K^{4/3}$. Thus the initial densities ($a_0,b_0$),
the diffusion constants ($D=D_a=D_b$), and the reaction rate $k$
determine all the parameters in $S(x,t)$ \cite{{galfi},{koza}}.

We turn now to the solution of eqn (\ref{move}).
In the absence of the source term, $S=0$, the globally 
stable homogeneous solutions are $m_h^0=\pm \sqrt{\epsilon/\gamma}$.
One can also find other homogeneous solutions $m(x,t)=m_h$ which, however, 
are unstable to small perturbations if 
$\vert m_h \vert <\sqrt{\epsilon/(3\gamma)}=m_s$.
The value $m_s$ gives the location of the spinodal line \cite{gun}. 
\begin{figure}[htb]
\centerline{
        \epsfysize=6cm
        \epsfbox{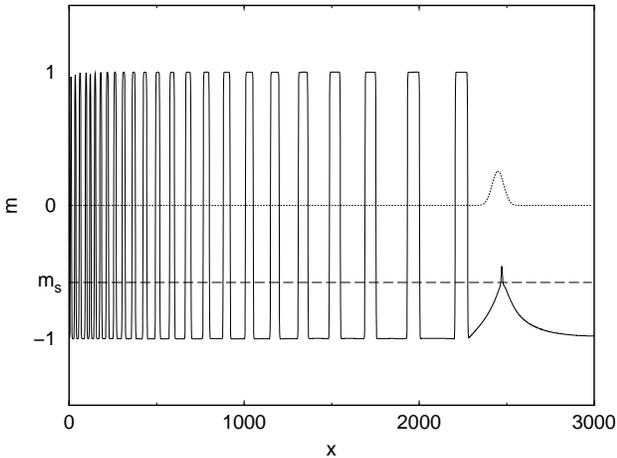}
           }
\vspace{0.5cm}
\caption{Magnetization profile obtained for the following values of 
the front parameters: $D_f=21.72$, $w_0=4.54$, and ${\cal A}=0.181$ with 
length, time, and the magnetization (concentration) 
measured in units of $\sqrt{\sigma\epsilon}$,
$\sigma/(\lambda \epsilon^2)$, and $\sqrt{\epsilon/\gamma}$, respectively.
The dotted line denotes the rate of local magnetization increase due to 
the source, $S$, measured in units of 
$\lambda\epsilon^{5/2}/(\gamma^{1/2}\sigma)$ and magnified by a factor
$2\cdot 10^5$. The dashed line is the magnetization at the 
spinodal line, $m_s=-1/\sqrt{3}$.}
\label{Fig2}
\end{figure}

In the presence of the source term $S$, equation~(\ref{move}) can be solved 
numerically using the initial condition 
$m(x,t)=-m_h^0$ and starting the source at the origin. 
The source alone would leave behind a uniform particle density
$c_0 \approx 0.85 K a_0 \sqrt{D/D_f}$~\cite{galfi} corresponding to a 
magnetization $m_0=2c_0-1$. The 
solution that evolves depends crucially on the value of $m_0$.

If $\vert m_0\vert>m_s$, the system is always outside of the unstable 
domain and a uniform band of precipitate is formed. Since there
is no noise in the system, this band is stable. 
Phase separation does take place when $\vert m_0\vert <m_s$ 
and a pattern similar to that 
shown in Fig.2 is observed. If $m_0$ is near the spinodal value 
then, at early stages of evolution, a nearly periodic 
set of narrow bands emerges that coarsens with time. 
Later on, the newly formed pattern crosses over to Liesegang type bands
with the distance between consecutive bands increasing.
If $m_0$ is such that the systems is deep in the unstable 
domain, well defined Liesegang bands form from the very beginning (Fig.2 shows 
an example of this type of evolution).

\begin{figure}[htb]
\centerline{
        \epsfysize=6cm
        \epsfbox{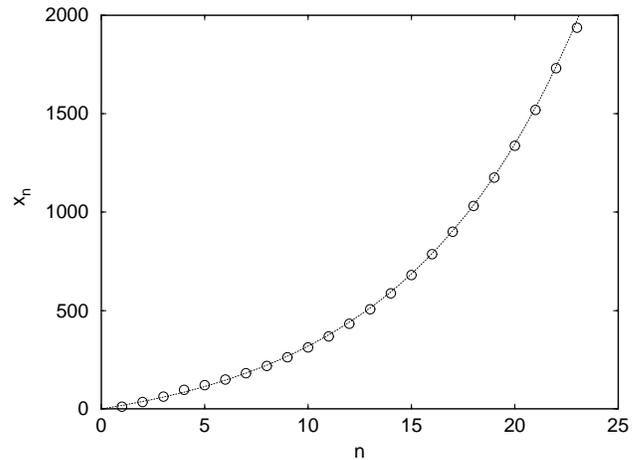}
           }
\vspace{0.5cm}
\caption{Positions of the bands $x_n$ as a function of their order 
of appearance $n$. Length is measured in units of $\sqrt{\sigma\epsilon}$.
The full line is the best fit given by equation~(\ref{fit}).}
\label{Fig3}
\end{figure}

In agreement with experiments, once formed, 
the Liesegang type bands are static on the time-scale 
we are able to observe (occasionally up to 100 bands were generated). 
Thus their spacing is well defined and
can be studied. Fig.3 displays the positions of the 
bands of the pattern exhibited in Fig.2. As one can see, 
a two-parameter fit of the form 
\be
x_n= Q [\exp({\tilde p}n)-1] \label{fit}
\ee
gives excellent agreement with the data (similar quality fits can be produced
for a wide range of parameter values).
Thus we see that the model contains the spacing law~(\ref{spacing}) 
with the spacing coefficient given by $1+p=\exp({\tilde p})$.

Next, one can investigate whether the patterns 
in our model obey the Matalon-Packter law. Calculating the $p$-s
for a set of $a_0$ at a fixed $b_0$, we find (see Fig.4) that the results 
indeed agree with the linear $b_0/a_0$ dependence given by eqn.(\ref{MatPac}).
The slopes and the intersections with the $p$ axis of the 
straight line fits provide us the functions $F(b_0)$ and $G(b_0)$. 
Due to the restricted interval of $b_0$ values that are 
available we can deduce only an approximate functional form for these
functions. It is clear, that both $F$ and $G$ are decreasing 
functions of their arguments, in agreement with experimental 
findings~\cite{{Matalon},{Packter}}. A fit to a power-law form gives
$F(b_0)\sim b_0^{-1.7}$ and $G(b_0)\sim b_0^{-1.1}$.

From the point of view of applications, it is important to know
the $a_0$ and $b_0$ dependence 
not only for $p$ but also for the amplitude $Q$ of the spacing law. 
In our model we can investigate this quantity as well
and our preliminary results indicate that $Q(a_0,b_0)\sim (a_0/b_0)^{0.4}$. 
The increase of $Q$ with $a_0$ is a somewhat surprising result. It means
that, as far as the number of bands in a given interval is concerned, 
there is a competition between $p$ and $Q$. When $a_0$ is increased, 
this number increases since $p\sim a_0^{-1}$ but it decreases due
to the change in the amplitude $Q\sim a_0^{0.4}$. This means that
in experiments where $n$ is finite (typically $n\sim 20-30$)
one could observe thinning of the bands as a result of the increase of
the concentration of the outer electrolyte.
 
\begin{figure}[htb]
\centerline{
        \epsfysize=6cm
        \epsfbox{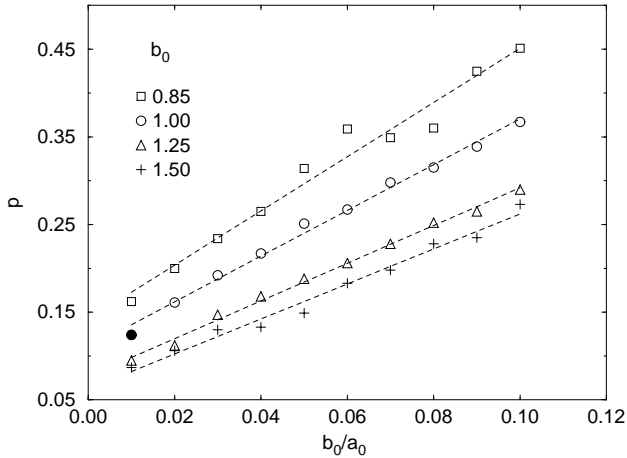}
           }
\vspace{0.5cm}
\caption{Spacing coefficient $p$ as a function of the ratio $b_0/a_0$ 
for several values of $b_0$. The unit of $b_0$ is fixed by the 
considering the front parameters given in Fig.2 and setting 
$b_0/a_0=0.01$ (filled circle in this figure).}
\label{Fig4}
\end{figure}

In conclusion we have proposed a new scenario for the formation of 
Liesegang patterns
based on a spinodal decomposition mechanism. Our approach has the advantage
of involving only a small number of parameters and 
there is no need to introduce artificial thresholds.
Our model yields the Matalon-Packter law and allows the calculation of both
the spacing coefficient $p(a_0,b_0)$ and the amplitude $Q(a_0,b_0)$ of the 
spacing law.

This simple scenario can be improved by studying the role of the noise 
either by adding it to 
the continuum model (\ref{move}) or by 
carrying out Monte-Carlo simulations on the kinetic Ising model with 
competing dynamics as described above. Work along these lines
is in progress~\cite{largepap}.

\section*{Acknowledgments}
We thank M. Zrinyi, P. Hantz, and T. Unger for useful discussions.
This work has been supported by the 
Swiss National Science Foundation and by the Hungarian Academy of Sciences 
(Grant No. OTKA T 029792).



\end{multicols}


\begin{references}
\bibitem{matpremark} Note that the Matalon-Packter law is usually 
                     written in the form $p=F(b_0)+G(b_0)/a_0$. We altered 
                     the definition of $G$ so that both $F$ and $G$ 
                     would be dimensionless.
\bibitem{spinodal} The spinodal line is defined in equilibrium  
                   mean-field theories as the line where system becomes unstable.
                   Otherwise it is a concept associated with the given
                   dynamics in the system: It is the line where 
                   the linearized dynamical equations become unstable. 
                   We shall use the latter definition.
\end{references}
\end{document}